\documentclass[twocolumn,showpacs,preprintnumbers]{revtex4}%
\usepackage{amssymb}
\usepackage{amsfonts}
\usepackage{amsmath}
\usepackage{graphicx}
\usepackage{dcolumn}
\usepackage{bm}%
\providecommand{\U}[1]{\protect\rule{.1in}{.1in}}
\begin{document}
\title{High intermodulation gain in a micromechanical Duffing resonator}
\author{R. Almog}
\email{almogr@tx.technion.ac.il}
\author{S. Zaitsev}
\author{O. Shtempluck }
\author{E. Buks}
\email{eyal@ee.technion.ac.il}
\affiliation{Department of Electrical Engineering, Technion, Haifa 32000 Israel}

\begin{abstract}
In this work we use a micromechanical resonator to experimentally study small
signal amplification near the onset of Duffing bistability. The device
consists of a PdAu beam serving as a micromechanical resonator excited by an
adjacent gate electrode. A large pump signal drives the resonator near the
onset of bistability, enabling amplification of small signals in a narrow
bandwidth. To first order, the amplification is inversely proportional to the
frequency difference between the pump and signal. We estimate the gain to be
about 15dB for our device.

\end{abstract}

\maketitle

Micro/Nanoelectromechanical resonators play a key role in microdevices for
applications such as sensing, switching, and filtering \cite{Roukes2000}%
,\cite{Craighead2000}. Understanding nonlinear dynamics in such devices is
highly important, both for applications and for basic research
\cite{Clelandbook}-\cite{Ayela}. The relatively small force needed for driving
a microresonator into the nonlinear regime is usually easily accessible,
enabling a variety of useful applications such as frequency mixing \cite{Erbe}
and frequency synchronization \cite{syncronization}. Since nano-scale
displacement detection is highly challenging, it is desirable to implement an
on-chip mechanical amplification mechanism. Previously, mechanical
amplification has been achieved using parametric amplification \cite{Rugar}%
,\cite{Carr}. Alternatively, amplification could be achieved by using a
bifurcating dynamical system \cite{byeb},\cite{Wiesenfeld}. In this work, we
employ nonlinear frequency mixing near the onset of Duffing bistability to
amplify small displacement signals. We demonstrate experimentally high signal
gain in this regime and compare with theoretical predictions.

\begin{figure}
[h]
\begin{center}
\includegraphics[
width=2.6671in
]%
{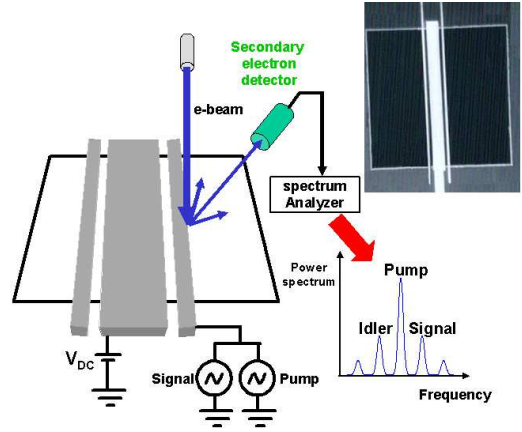}%
\caption{The experimental setup. The inset shows an electron micrograph of the
device, consisting of two suspended doubly clamped micromechanical resonators.
Each resonator is of length $l$=100$\mu$m, width $w$=0.6$\mu$m, and thickness
$t$=0.25$\mu$m, centered around a gate electrode with $d=4\mu m$ gap. The
device is mounted inside an SEM operated in a spot mode to detect the
resonator's displacement. The displacement signal is probed by a secondary
electron detector and measured using a spectrum analyzer. }%
\label{SetupAndDevice}%
\end{center}
\end{figure}

The device under study is a mechanical resonator consisting of a suspended
doubly clamped PdAu beam, located adjacent to a static gate electrode. An
electron micrograph of the device is shown in the inset of Fig.
\ref{SetupAndDevice}, and its dimensions are given in the figure's caption.%

The nonlinear dynamics of the fundamental mode of a doubly clamped beam driven
by an external force per unit mass $F(t)$ can be described by a Duffing
oscillator equation for a single degree of freedom $x$ \cite{Nayfebook}%
\begin{equation}
\ddot{x}+2\mu\dot{x}+\omega_{0}^{2}(1+\kappa x^{2})x=F(t), \label{eom}%
\end{equation}

where $\mu$ is the damping constant, $\omega_{0}/2\pi$ is the resonance
frequency of the fundamental mode, and $\kappa$ is the cubic nonlinear
constant. For small displacement amplitudes, the nonlinearity is originated by
beam's mid-plane stretching due to the immovable boundaries, which tend to
harden the beam ($\kappa>0)$. For higher amplitudes, however, the contribution
of the applied electric force, which tend to soften the beam, becomes dominant.

Generally, for resonators driven using a bias voltage applied to a side
electrode, Eq. \ref{eom} should contain additional parametric terms
\cite{Rugar},\cite{Blencowe}. In our case however, the prefactors of these
parametric terms is at least one order smaller below threshold, thus negligible.

The device has a quality factor of $Q=\omega_{0}/2\mu\approx2000$ (at $10^{-5}%
\operatorname{torr}%
)$ and the fundamental mode resonance frequency $\omega_{0}/2\pi$ is in the
range of 560 $%
\operatorname{kHz}%
$ to 630$%
\operatorname{kHz}%
.$

To investigate nonlinear amplification, the resonator is driven by an applied
force $F(t)=f_{p}\cos(\omega_{p}t)+f_{s}\cos(\omega_{s}t+\varphi),$ composed
of an intense \textit{pump} with frequency $\omega_{p}=\omega_{0}+\sigma,$
amplitude $f_{p},$ and a small force (called \textit{signal}) with frequency
$\omega_{s}=\omega_{p}+\delta,$ relative phase $\varphi$, and amplitude
$f_{s}$, where $f_{s}<<f_{p}$ and $\sigma,\delta<<\omega_{0}.$ This is
achieved by applying a voltage of the form $V=V_{dc}+v_{p}\cos(\omega
_{p}t)+v_{s}\cos(\omega_{s}t+\varphi)$ where $V_{dc}$ is a dc bias and
$v_{s}<<v_{p}<<V_{dc}$. The resonator's displacement has spectral components
at $\omega_{p},$ $\omega_{s},$ and at the intermodulations $\omega_{p}\pm
k\delta$ where $k$ is an integer. The one at frequency $\omega_{i}=\omega
_{p}-\delta$ is called the idler component, as in nonlinear optics.

\begin{figure}
[h]
\begin{center}
\includegraphics[
width=3.0372in
]%
{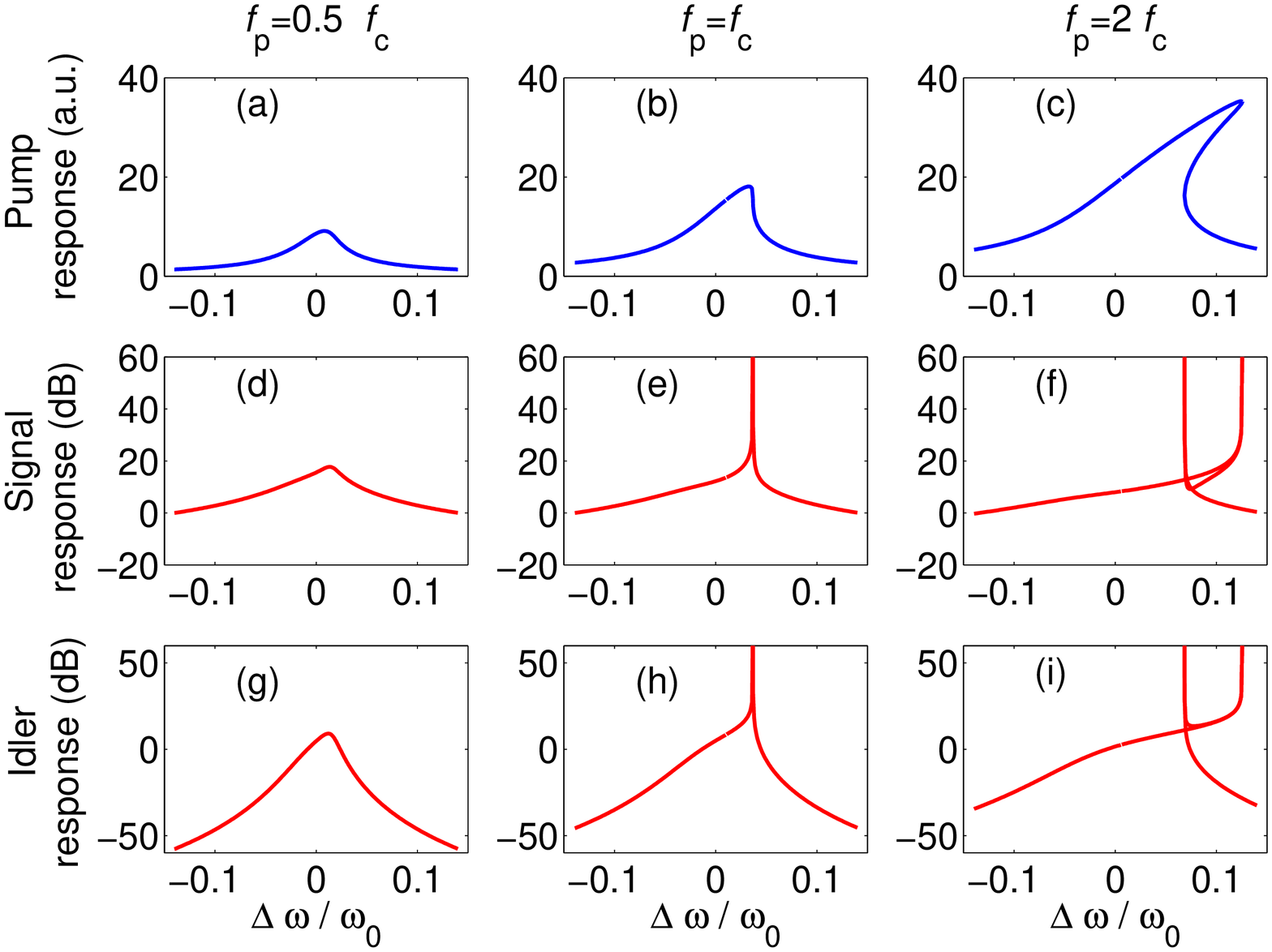}%
\caption{Calculation of the pump, signal, and idler responses ($\left\vert
a_{p}\right\vert ,$ $\left\vert a_{s}\right\vert ,$ $\left\vert a_{i}%
\right\vert $) for vanishing offset frequency $\delta$, shown for sub-critical
case $f_{p}=0.5f_{c}$ (a,d,g)$,$ critical case $f_{p}=f_{c}$ (b,e,h), and
over-critical case $f_{p}=2f_{c}$ (c,f,i)$.$ The y-axis of the pump is shown
in a linear scale while the signal's and idler's response are normalized to
the signal's excitation amplitude and are shown in a logarithmic scale. The
signal's and idler's response diverge at the critical point and at the jump
points. The parameters for this example are $\kappa=10^{-4}\operatorname{m}%
^{-2},$ $\mu=10^{2}\operatorname{Hz}$, $\omega_{0}/2\pi=1\operatorname{MHz},$
and $\delta/2\pi=10\operatorname{Hz}.$}%
\label{calc_crv}%
\end{center}
\end{figure}

In the slowly varying envelope method \cite{Nayfebook}, the displacement $x$
is written as%
\begin{equation}
x(t)=\frac{1}{2}A(t)e^{i\omega_{p}t}+c.c.\text{,} \label{xt}%
\end{equation}
where $A(t)$ is a slowly varying function (relative to the time scale
$1/\omega_{p}).$ Substituting Eq. (\ref{xt}) in the equation of motion
(\ref{eom}) and neglecting the $d^{2}A/dt^{2}$ term yields%
\begin{equation}
\frac{dA}{dt}=-(\dfrac{\omega_{0}}{2Q}+i\sigma)A+i\dfrac{3}{8}\kappa\omega
_{0}A^{2}A^{\ast}+\dfrac{1}{2i\omega_{0}}(f_{p}+f_{s}e^{i(\delta t+\varphi
)})\text{.} \label{dAdt}%
\end{equation}
$A(t)$ can be written as%
\begin{equation}
A(t)=a_{p}+a_{s}e^{i\delta t}+a_{i}e^{-i\delta t}\text{ },\text{ } \label{At}%
\end{equation}
where the complex numbers $a_{p,}$ $a_{s}$ and $a_{i}$ are the pump, signal
and idler components of $A(t)$ respectively, and $\left\vert a_{s}\right\vert
,\left\vert a_{i}\right\vert <<\left\vert a_{p}\right\vert . $ Substituting
Eq. (\ref{At}) in Eq. (\ref{dAdt}) and keeping small terms up to first order,
leads to \cite{byeb}
\begin{subequations}
\label{eq15}%
\begin{align}
a_{s}  &  =\dfrac{\dfrac{1}{2\omega_{0}}f_{s}e^{i\varphi}-\dfrac{3}{8}%
\kappa\omega_{0}a_{p}^{2}a_{i}^{\ast}}{\dfrac{3}{4}\kappa\omega_{0}|a_{p}%
|^{2}-\delta-\sigma+i\frac{\omega_{0}}{2Q}},\\
a_{i}  &  =\dfrac{-\dfrac{3}{8}\kappa\omega_{0}a_{p}^{2}a_{s}^{\ast}}%
{\dfrac{3}{4}\kappa\omega_{0}|a_{p}|^{2}-\delta-\sigma+i\frac{\omega_{0}}{2Q}%
}.
\end{align}

The pump response $\left\vert a_{p}\right\vert $ in the absent of any
additional signal is shown in Fig. \ref{calc_crv}, panels (a),(b) and (c)
\cite{byeb}. Above some critical driving amplitude $f_{c}$, the response
becomes a multi-valued function of the frequency in some finite frequency
range, and the system becomes bistable with jump points in the frequency
response. We refer to the onset point of bistability (which is also a
saddle-node bifurcation point) as the critical point. When the pump is tuned
to the critical point ($\sigma=\sqrt{3}\omega_{0}/2Q,$ $|a_{p}|^{2}%
=8/3\sqrt{3}\kappa Q)$ \cite{LL} and $\delta\rightarrow0,$ we expect high
amplification of both signal and idler. In this limit
\end{subequations}
\begin{equation}
\left\vert a_{s}\right\vert \simeq\left\vert a_{i}\right\vert \simeq
\frac{f_{s}}{2\omega_{0}\delta}\text{ .}%
\end{equation}
Thus, in our model which assumes that $\left\vert a_{s}\right\vert $ and
$\left\vert a_{i}\right\vert $ are small, and takes nonlinearity into account
only to lowest order, the amplification diverges in the limit $\delta
\rightarrow0$. When $\left\vert a_{s}\right\vert $ and $\left\vert
a_{i}\right\vert $ become comparable with $\left\vert a_{p}\right\vert $,
however, the former assumptions are no longer valid and higher order terms
have to be taken into account.

The pump, signal, and idler's responses were calculated analytically
\cite{byeb} and are shown in Fig. \ref{calc_crv}. For a small $f_{p}$, the
signal's response is nearly Lorentzian, while for $f_{p}>f_{c}$, both signal
and idler response diverge near the jump points.%

The resonators are fabricated using bulk-nano-machining process together with
electron beam lithography \cite{ebmlr}. The experimental setup is shown in
Fig. \ref{SetupAndDevice}. Measurement of mechanical vibration is done at room
temperature,\textit{\ in-situ} a scanning electron microscope (SEM)\ where the
imaging system of the microscope is employed for displacement detection
\cite{ebmlr}. The three spectral components $\omega_{p},$ $\omega_{s},$ and
$\omega_{i}$ of the displacement are measured using a spectrum analyzer.%

\begin{figure}
[h]
\begin{center}
\includegraphics[
width=3.058in
]%
{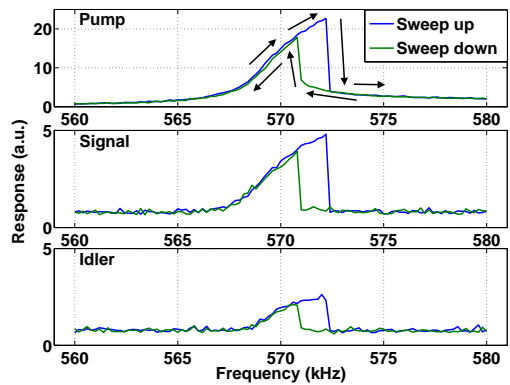}%
\caption{ Simultaneous measurement of the pump, signal and idler spectral
components of the mechanical displacement. The excitation frequency is swept
upward (blue line) and downward (green line). The arrows in the pump's plot
indicate the hysteresis loop. The excitation parameters are: pump ac voltage
$v_{p}=0.5\operatorname{V},$ $v_{p}/v_{s}=6,$ frequency offset $\delta
/2\pi=1\operatorname{kHz}$ and $V_{dc}=5\operatorname{V}.$ The horizontal axis
is the pump frequency for all three plots. The pump signal and idler exhibit
simultaneous jumps, as expected. }%
\label{hysteresis}%
\end{center}
\end{figure}
\begin{figure}
[h]
\begin{center}
\includegraphics[
width=3.0234in
]%
{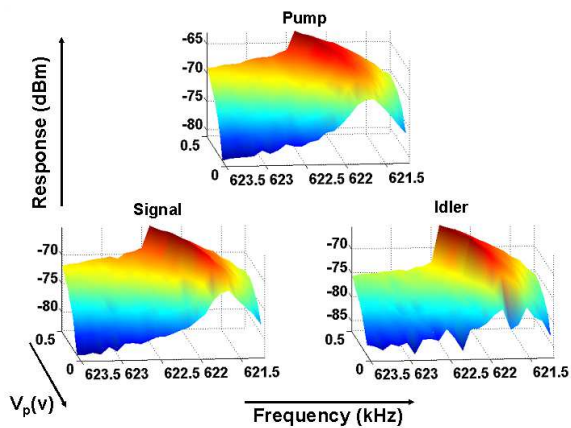}%
\caption{Mesh plots showing the response of the pump, signal and idler. The
horizontal axis is the pump's frequency $\omega_{p}$, the diagonal axis is the
pump's ac voltage $v_{p}$ and the vertical axis is the response (displacement)
axis in logarithmic scale. For each frequency, $v_{p}$ is scanned from 0 to
0.5$\operatorname{V}$, $v_{p}/v_{s}=6,$ $\delta/2\pi=100\operatorname{Hz}%
,\ V_{dc}=5\operatorname{V}.$ Note that the pump's response undergoes a jump
along a line in the ($v_{p},\omega_{p})$ plane, starting from the bifurcation
point. Along the same line, the spectral components of \ the signal and idler
obtain their maximum value.}%
\label{IMJ}%
\end{center}
\end{figure}
A typical mechanical response is shown in fig. \ref{hysteresis}. The pump's
frequency is swept upward and then back downward. As expected, we find
hysteretic response and simultaneous jumps for the pump, signal, and idler
spectral components. In Fig. \ref{IMJ} the mechanical responses of the pump,
signal and idler are depicted as a function of the pump's frequency
$\omega_{p}/2\pi$ and the pump's ac voltage $v_{p}$. For each frequency, the
voltage $v_{p}$ is scanned from $0$ to $0.5%
\operatorname{V}%
$. The results show a good agreement with theory. As expected, we observe high
signal amplification near the jump points. The amplification can be quantified
in logarithmic scale as%
\begin{equation}
G\equiv20\log(\left\vert \frac{a_{s,pump\_on}}{a_{s,pump\_off}}\right\vert
)\text{ .}%
\end{equation}
The highest value of $G$, obtained near one of the jump points is $15$dB.
Note, however, that this value is an underestimation of the actual gain due to
nonlinearity of our displacement detection scheme.

In conclusion, we have shown that a Duffing micromechanical resonator, driven
into the bistability regime, can be employed as a high gain narrow band
mechanical amplifier.

This work is supported by Intel Corp. and by the Israel-US\ Binational Science
Foundation (BSF\ grant no. 2004362). The authors wish to thank Bernard Yurke
for helpful discussions.

\end{document}